\documentclass[prb,twocolumn,letterpaper,superscriptaddress]{revtex4-1}
\usepackage[space]{grffile}
\usepackage{bm,graphicx,graphics,amsmath,amssymb,bm,epsfig,color}
\usepackage{euscript,tabularx}
\usepackage{longtable}
\usepackage{float}
\usepackage[draft]{todonotes}

\begin{document}

\title{Generation of coherent spin-wave modes in Yttrium Iron Garnet
  microdiscs\\ by spin-orbit torque}

\author{M. Collet}
\affiliation{Unit\'e Mixte de Physique CNRS/Thales and Universit\'e
  Paris Sud 11, 1 av. Fresnel, 91767 Palaiseau, France}

\author{X. de Milly}
\affiliation{Service de Physique de l'\'Etat Condens\'e (CNRS UMR
  3680), CEA Saclay, 91191 Gif-sur-Yvette, France}

\author{O. d'Allivy Kelly}
\affiliation{Unit\'e Mixte de Physique CNRS/Thales and Universit\'e
  Paris Sud 11, 1 av. Fresnel, 91767 Palaiseau, France}

\author{V. V. Naletov} 
\affiliation{INAC-SPINTEC, CEA/CNRS and Univ. Grenoble Alpes, 38000
  Grenoble, France}
\affiliation{Institute of Physics, Kazan Federal University, Kazan
    420008, Russian Federation}

\author{R. Bernard} 
\affiliation{Unit\'e Mixte de Physique CNRS/Thales
  and Universit\'e Paris Sud 11, 1 av. Fresnel, 91767 Palaiseau,
  France}

\author{P. Bortolotti} 
\affiliation{Unit\'e Mixte de Physique CNRS/Thales
  and Universit\'e Paris Sud 11, 1 av. Fresnel, 91767 Palaiseau,
  France}

\author{V. E. Demidov}
\affiliation{Department of Physics, University of Muenster, 48149 Muenster, Germany}

\author{S. O. Demokritov} 
\affiliation{Department of Physics, University of Muenster, 48149 Muenster, Germany}
\affiliation{Institute of Metal Physics, Ural Division of RAS,
  Yekaterinburg 620041, Russian Federation}

\author{J. L. Prieto} 
\affiliation{Instituto de Sistemas Optoelectr\'onicos y
  Microtecnolog\'{\i}a (UPM), Madrid 28040, Spain}

\author{M. Mu\~noz}
\affiliation{Instituto de Microelectr\'onica de Madrid (CNM, CSIC),
  Madrid 28760, Spain}

\author{V. Cros}
\affiliation{Unit\'e Mixte de Physique CNRS/Thales and Universit\'e
  Paris Sud 11, 1 av. Fresnel, 91767 Palaiseau, France}

\author{A. Anane}
\affiliation{Unit\'e Mixte de Physique CNRS/Thales and Universit\'e
  Paris Sud 11, 1 av. Fresnel, 91767 Palaiseau, France}

\author{G. de Loubens} 
\affiliation{Service de Physique de l'\'Etat Condens\'e (CNRS UMR
  3680), CEA Saclay, 91191 Gif-sur-Yvette, France}
\email[Corresponding author:]{ gregoire.deloubens@cea.fr}

\author{O. Klein} 
\affiliation{INAC-SPINTEC, CEA/CNRS and Univ. Grenoble Alpes, 38000
  Grenoble, France}
\email[Corresponding author:]{ olivier.klein@cea.fr}

\date{\today}

\begin{abstract}

  \textbf{Spin-orbit effects
    \cite{manchon09,miron11,rojas13a,mellnik14} have the potential of
    radically changing the field of spintronics by allowing transfer
    of spin angular momentum to a whole new class of materials. In a
    seminal letter to Nature \cite{kajiwara10}, Kajiwara \textit{et
      al.} showed that by depositing Platinum (Pt, a normal metal) on
    top of a 1.3~$\mu$m thick Yttrium Iron Garnet (YIG, a magnetic
    insulator), one could effectively transfer spin angular momentum
    through the interface between these two different materials. The
    outstanding feature was the detection of auto-oscillation of the
    YIG when enough dc current was passed in the Pt. This finding has
    created a great excitement in the community for two reasons:
    first, one could control electronically the damping of insulators,
    which can offer improved properties compared to metals, and here
    YIG has the lowest damping known in nature; second, the damping
    compensation could be achieved on very large objects, a
    particularly relevant point for the field of magnonics
    \cite{kruglyak10a,serga10} whose aim is to use spin-waves as
    carriers of information. However, the degree of coherence of the
    observed auto-oscillations has not been addressed in
    ref. \cite{kajiwara10}. In this work, we emphasize the key role of
    quasi-degenerate spin-wave modes, which increase the threshold
    current. This requires to reduce both the thickness and lateral
    size in order to reach full damping compensation
    \cite{hamadeh14b}, and we show clear evidence of coherent
    spin-orbit torque induced auto-oscillation in micron-sized YIG
    discs of thickness 20~nm.}

\end{abstract}

\maketitle

% ****************************************************************
% Introduction
When spin transfer effects were first introduced by Slonczweski and
Berger in 1996 \cite{slonczewski96,berger96}, the authors immediately
recognized that the striking signature of the transfer process would
be the emission of microwave radiation when the system is pumped out
of equilibrium by a dc current. Since the spin transfer torque on the
magnetisation is collinear to the damping torque, there is an
instability threshold when the natural damping is fully compensated by
the external flow of angular momentum, leading to spin-wave
amplification through stimulated emission. Using analogy to light, the
effect was called SWASER \cite{berger96}, where SW stands for
spin-wave. Until 2010, all SWASER devices required a charge current
perpendicular to the plane to transfer angular momentum between
different magnetic layers \cite{slonczewski96,berger96}. This implied
that the effect was restricted to conducting materials. The situation
has radically changed since spin-orbit effects such as the spin Hall
effect (SHE) \cite{dyakonov71, hirsch99} are used to produce spin
currents in normal metals. Here a right hand side rule links the
deflected direction of the electron and the orientation of its
spin. This allows the creation of a pure spin current transversely to
the charge current, with an efficiency given by the spin Hall angle
$\Theta_\text{SH}$. Using a metal with large $\Theta_\text{SH}$, such
as Pt, a charge current flowing in plane generates a pure spin current
flowing perpendicular to the plane, which can eventually be
transferred through an interface with ferromagnetic metals, resulting
in the coherent emission of spin-waves \cite{demidov12}, but also with
non-metals such as YIG \cite{kajiwara10}.

The microscopic mechanisms of transfer of angular momentum between a
normal metal and a ferromagnetic layer are quite different depending
on the latter being metallic or not. In the first case, electrons in
each layer have the possibility to penetrate the other one, whereas in
the second case the transfer takes place exactly and solely at the
interface. It is thus much more sensitive to the imperfection of the
interface. Still, a direct experimental evidence that spin current can
indeed cross such an hybrid interface is through the so-called spin
pumping effect \cite{tserkovnyak05}: adding a normal metal on top of
YIG increases its ferromagnetic resonance (FMR) linewidth
\cite{heinrich11}, which is due to the new relaxation channel at the
interface through which angular momentum can escape and get absorbed
in the metal. This effect being interfacial, the broadening scales as
$1/t_\text{YIG}$, where $t_\text{YIG}$ is the thickness of YIG. Even
for YIG, whose natural linewidth is only a few Oersted at 10~GHz, it
is hardly observable if $t_\text{YIG}$ exceeds a couple hundreds of
nanometers. For these thick films though, the spin pumping can still
be detected through inverse spin Hall effect (ISHE). In a normal metal
with strong spin-orbit interaction, the pumped spin current is
converted into a transverse charge current. This generates a voltage
proportional to the length of the sample across the metal, which can
easily reach several tens of microvolts in millimeter-sized
samples. Since the first experiment by Kajiwara \textit{et al.}
\cite{kajiwara10}, many studies reported the ISHE detection of FMR
using different metals on YIG layers \cite{hahn13,mendes14,wang14b},
hereby providing clear evidence of at least partial transparency of
the hybrid YIG$|$metal interfaces to spin currents.  Due to Onsager
relations, these results made the community confident that a spin
current could thus be injected from metals to YIG and lead to the
SWASER effect.

From the beginning it was anticipated that the key to observe
auto-oscillations in non-metals was to reduce the threshold
current. The first venue is of course to choose a material whose
natural damping is very low. In this respect YIG is the optimal
choice. The second thing is to reduce the thickness since the
spin-orbit torque (SOT) is an interfacial effect. This triggered an
effort in the fabrication of ultra-thin films of YIG of very high
dynamical quality \cite{sun12,kelly13}. For 20~nm thick YIG films with
damping constant as low as $\alpha = 2.3\cdot10^{-4}$, a striking
result was that there were no evidence of auto-oscillations in
millimeter-sized samples at the highest dc current possible in the top
Pt layer \cite{kelly13}, \textit{i.e.}, before it evaporates. It is
worth mentioning at this point that reducing further the thickness or
the damping parameter of such ultra-thin YIG films \cite{chang14} does
not help anymore in decreasing the threshold current, as the relevant
value of the damping is that of the YIG$|$Pt hybrid, which ends up to
be completely dominated by the spin-pumping contribution.

But most notably, none of these high-quality ultra-thin YIG films
display a purely homogeneous FMR line. The reason for that is well
known. In such extended films, there are many degenerate modes with
the main, uniform FMR mode, which through the process of two-magnon
scattering broaden the linewidth \cite{arias99,mcmichael04}. A
striking evidence of these degenerate modes can be obtained by
parametrically pumping the SW modes. It reveals an uncountable number
of modes which are at the same energy as the FMR mode
\cite{hahn13a}. Any threshold instability will be affected by the
presence of those modes, as learnt from LASERs where mode competition
is known to have a strong influence on the emission threshold
\cite{asryan96}. Thus, the next natural step was to reduce as well the
lateral size in order to lift the degeneracy between modes through
confinement. The first microstructures of YIG appeared revealing that
the patterning indeed narrowed the linewidth through a decrease of the
inhomogeneous part \cite{hahn14}. The effect is clear in the
perpendicular geometry, where magnon-magnon processes are suppressed
owing to the fact that the FMR mode lies at the bottom of the SW
dispersion relation. However, this is not the case in the parallel
geometry where the FMR mode is not the lowest energy SW mode. Even
then, we showed that the linewidth in a micron-sized YIG$|$Pt disc
could be still tuned thanks to SOT \cite{hamadeh14b}. In the
following, we describe the direct electrical detection of
auto-oscillations in similar samples and show that the threshold
current is increased by the presence of quasi-degenerate SW modes.

% ****************************************************************
% Straight to the main results

\begin{figure}
  \includegraphics[width=7.5cm]{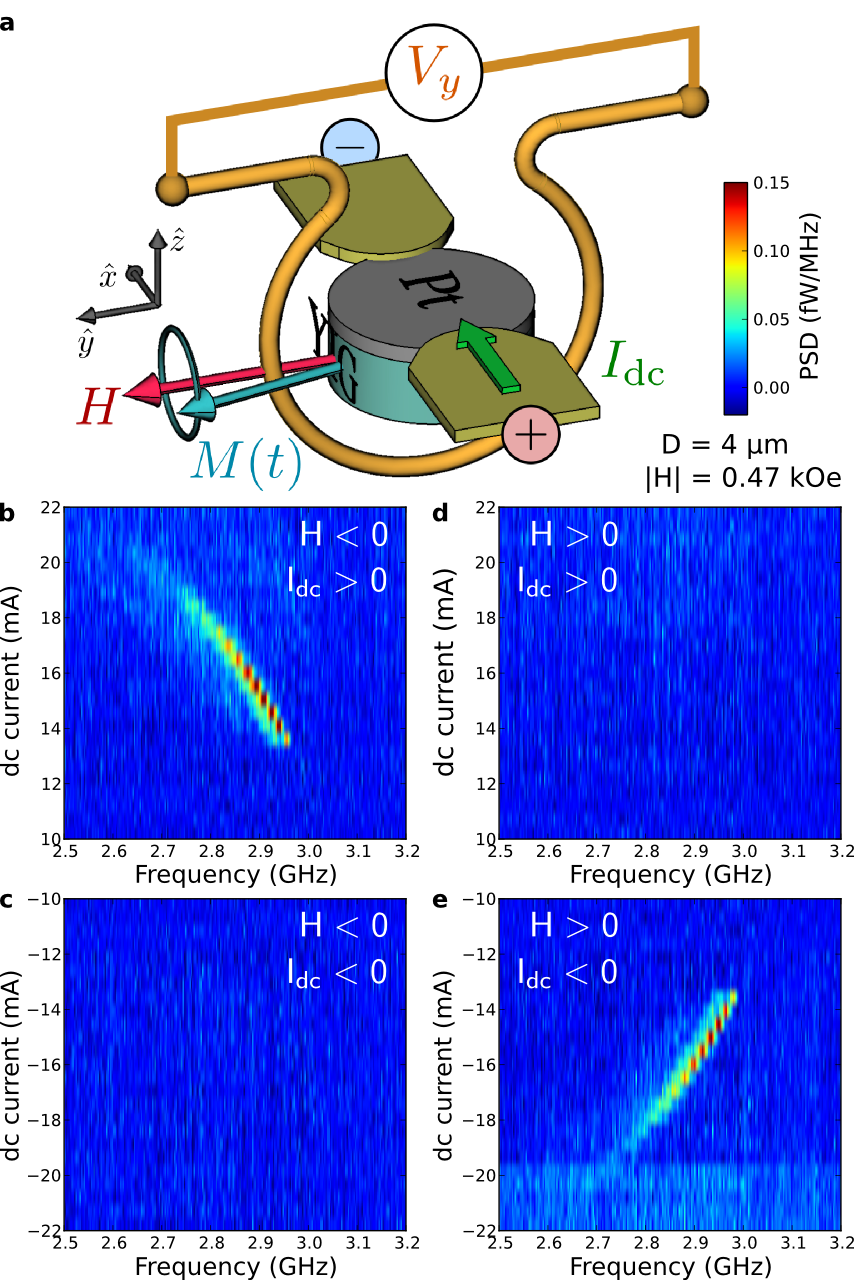}
  \caption{\textbf{Inductive detection of auto-oscillations in a
      YIG$|$Pt microdisc.} \textbf{a}, Sketch of the sample and
    measurement configuration. The bias field $H$ is oriented
    transversely to the dc current $I_\text{dc}$ flowing in Pt. The
    inductive voltage $V_y$ produced in the antenna by the precession
    of the YIG magnetisation is amplified and monitored by a spectrum
    analyser. \textbf{b--e}, Power spectral density maps measured at
    fixed $|H|=0.47$~kOe and variable $I_\text{dc}$. The four
    quadrants correspond to different possible polarities of $H$ and
    $I_\text{dc}$. }
  \label{FIG1}
\end{figure}

We study magnetic microdiscs with diameter 2~$\mu$m and 4~$\mu$m which
are fabricated based on a hybrid YIG(20~nm)$|$Pt(8~nm) bilayer. The
20~nm thick YIG layer is grown by pulsed laser deposition
\cite{kelly13} and the 8~nm thick Pt layer is sputtered on top of it
\cite{rojas14}. Their physical parameters are summarized in Table
I. We stress that the extended YIG film is characterized by a low
Gilbert damping parameter $\alpha_0 = (4.8 \pm 0.5) \cdot 10^{-4}$ and
a remarkably small inhomogeneous contribution to the linewidth,
$\Delta H_0 = 1.1 \pm 0.3$~Oe. Each microdisc is connected to
electrodes enabling the injection of a dc current $I_\text{dc}$ in the
Pt layer, and a microwave antenna is defined around it to obtain an
inductive coupling with the YIG magnetisation, as shown schematically
in Fig. 1a.

\begin{table*}
  \caption{Physical parameters of the Pt and bare YIG layers, and of the hybrid YIG$|$Pt bilayer.}
  \begin{ruledtabular}
    \begin{tabular}{c | c c c c }
      Pt & $t_\text{Pt}$ (nm) & $\rho$ ($\mu\Omega$.cm) & $\lambda_\text{SD}$ (nm) & $\Theta_\text{SH}$ \\ \hline 
      from ref.\cite{rojas14}& 8 & $17.3 \pm 0.6$ & $3.4 \pm 0.4$ & $0.056 \pm 0.010$\\ \hline \hline 
      YIG & $t_\text{YIG}$ (nm) & $\alpha_0$ & $4 \pi M_s$ (G) & $\gamma$ ($10^7$~rad.s$^{-1}$.G$^{-1}$) \\ \hline
      this study & 20 & $(4.8 \pm 0.5) \cdot 10^{-4}$ & $2150 \pm 50$ & $1.770 \pm 0.005$ \\ \hline \hline 
      YIG$|$Pt & $t_\text{YIG}|t_\text{Pt}$ (nm) & $\alpha$ & $g_{\uparrow \downarrow}$ ($ 10^{18}$~m$^{-2}$) & $T$  \\ \hline 
      this study & 20$|$8 & $(2.05 \pm 0.1) \cdot 10^{-3}$ &  $3.6 \pm 0.5$  & $0.2 \pm 0.05$
    \end{tabular}
  \end{ruledtabular}
  \label{tab:mat}
\end{table*}

\begin{figure*}
  \includegraphics[width=17.5cm]{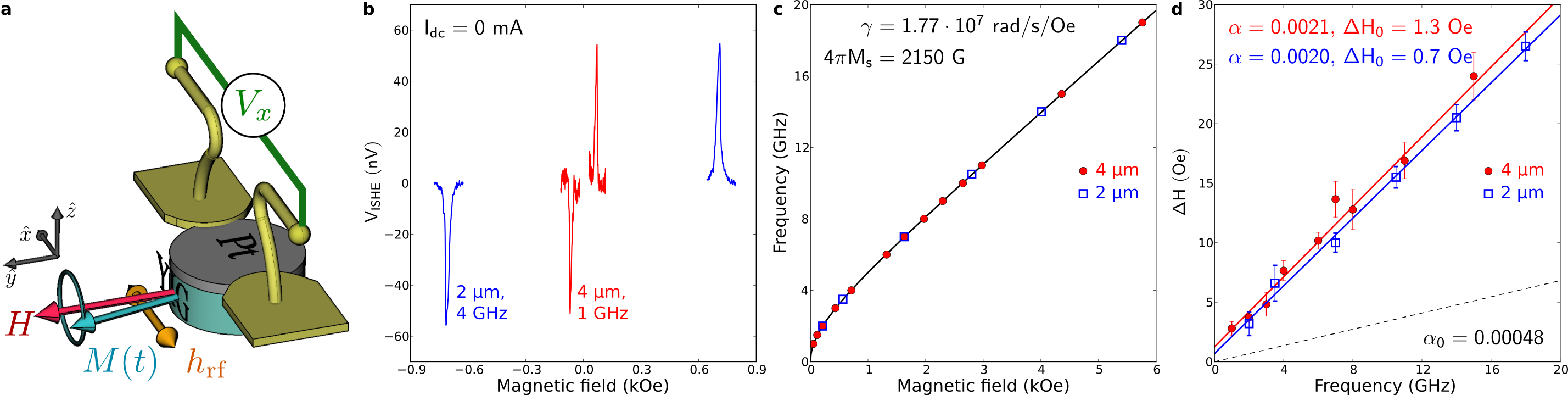}
  \caption{\textbf{ISHE-detected FMR spectroscopy in YIG$|$Pt
      microdiscs.}  \textbf{a}, Sketch of the sample and measurement
    configuration. The bias field $H$ is oriented perpendicularly to
    the Pt electrode and to the excitation field $h_\text{rf}$
    produced by the antenna at fixed microwave frequency. The dc
    voltage $V_x$ across Pt is monitored as a function of the magnetic
    field. \textbf{b}, ISHE-detected FMR spectra of the 4~$\mu$m and
    2~$\mu$m YIG(20~nm)$|$Pt(8~nm) discs at 1~GHz and 4~GHz,
    respectively. \textbf{c}, Dispersion relation of the main FMR mode
    of the microdiscs. The continuous line is a fit to the Kittel law.
    \textbf{d}, Frequency dependence of the FMR linewidth in the two
    microdiscs. The vertical bars show the mean squared error of the
    lorentzian fits. The continuous lines are linear fits to the
    data. The dashed line shows the homogeneous contribution of the
    bare YIG.}
  \label{FIG2}
\end{figure*}

First, we monitor with a spectrum analyser the voltage produced in the
antenna by potential auto-oscillations of the 4 $\mu$m YIG disc as a
function of the dc current $I_\text{dc}$ injected in Pt. The in-plane
magnetic field $H=0.47$~kOe is applied in a transverse direction with
respect to $I_\text{dc}$, as shown in Fig. 1a. This is the most
favorable configuration to compensate the damping and obtain
auto-oscillations in YIG, as spins accumulated at the YIG$|$Pt
interface due to SHE in Pt will be collinear to its
magnetisation. Color plots of the inductive signal measured as a
function of the relative polarities of $H$ and $I_\text{dc}$ are
presented in Fig. 1b--e. At $H<0$, we observe in the power spectral
density (PSD) a peak which starts at around 2.95~GHz and $13$~mA and
then shifts towards lower frequency as $I_\text{dc}$ is increased
(Fig. 1b), a clear signature that spin transfer occurs through the
YIG$|$Pt interface. The linewidth of the emission peak, lying in the
10--20~MHz range for $13<I_\text{dc}<17$~mA, also proves the coherent
nature of the detected signal. An identical behaviour is observed at
$H>0$ and $I_\text{dc}<0$ (Fig. 1d). In contrast, the PSD remains flat
in the two other cases (Fig. 1c--d). Therefore, an auto-oscillation
signal is detected only if $H \cdot I_\text{dc} < 0$, in agreement
with the expected symmetry of SHE.

% ****************************************************************
% FMR characterization (Vishe, spin pumping)

In order to characterize the flow of angular momentum across the
YIG$|$Pt interface, we now perform ISHE-detected FMR spectroscopy on
our microdiscs. The configuration of this experiment is similar to the
previous case, but now the antenna generates a uniform microwave field
$h_\text{rf}$ to excite the FMR of YIG while the dc voltage across Pt
is monitored at zero current (see Fig. 2a). In other words, we perform
the reciprocal experiment of the one presented in Fig. 1. As described
in the introduction, a voltage $V_\text{ISHE}$ develops across Pt when
the FMR conditions are met in YIG. This voltage changes sign as the
field is reversed, which is expected from the symmetry of ISHE, and
shown in Fig. 2b, where the FMR spectra of the 4~$\mu$m and 2~$\mu$m
microdiscs are respectively detected at 1~GHz and 4~GHz. We also note
that for a given field polarity, the product between $V_\text{ISHE}$
and $I_\text{dc}$ must be negative to compensate the damping
\cite{hamadeh14b}, which enables to observe auto-oscillations in
Fig. 1.

From these ISHE measurements, the dispersion relation and frequency
dependence of the full linewidth at half maximum of the main FMR mode
can be determined, as shown in Figs. 2c and 2d, respectively. The
dispersion relation follows the expected Kittel law. The damping
parameters of the 4~$\mu$m and 2~$\mu$m microdiscs, extracted from
linear fits to the data, $\Delta H = 2\alpha\omega/\gamma+\Delta H_0$
(continuous lines in Fig. 2d, $\omega$ is the pulsation frequency and
$\gamma$ the gyromagnetic ratio), are found to be similar with an
average value of $\alpha = (2.05 \pm 0.1) \cdot 10^{-3}$. The small
inhomogeneous contribution to the linewidth observed in both
microdiscs, $\Delta H_0 = 1.3 \pm 0.4$~Oe and $\Delta H_0 = 0.7 \pm
0.4$~Oe, respectively, decreases with the diameter and is attributed
to the presence of several unresolved modes within the resonance line
\cite{hamadeh14b}. In order to emphasize the increase of damping due
to Pt, we have reported in Fig. 2d the broadening produced by the
homogeneous contribution of the bare YIG using a dashed line. The
observed increase of damping is due to spin pumping
\cite{tserkovnyak05,heinrich11},
\begin{equation}
  \alpha-\alpha_0 = g_{\uparrow\downarrow} \frac{\gamma \hbar}{4\pi M_st_\text{YIG}} \, ,
  \label{eq:alpha_sp}
\end{equation}
where $\hbar$ is the reduced Planck constant, $M_s$ the saturation
magnetisation and $g_{\uparrow \downarrow}$ the spin-mixing
conductance of the YIG$|$Pt interface. This allows us to extract
$g_{\uparrow \downarrow} = (3.6 \pm 0.5) \cdot 10^{18}$~m$^{-2}$,
which lies in the same window as previously reported values
\cite{jungfleisch13a,hahn14}. From the spin-mixing conductance
$g_{\uparrow \downarrow}$, the spin diffusion length
$\lambda_\text{SD}$ and the resistivity $\rho$ of the Pt layer, we can
also calculate the transparency of the YIG$|$Pt interface to spin
current \cite{chen13}, $T=0.2 \pm 0.05$. The physical parameters
extracted for the YIG$|$Pt hybrid bilayer are summarized in the last
raw of Table I.

% ****************************************************************
% Angle dependence

\begin{figure}
  \includegraphics[width=8.5cm]{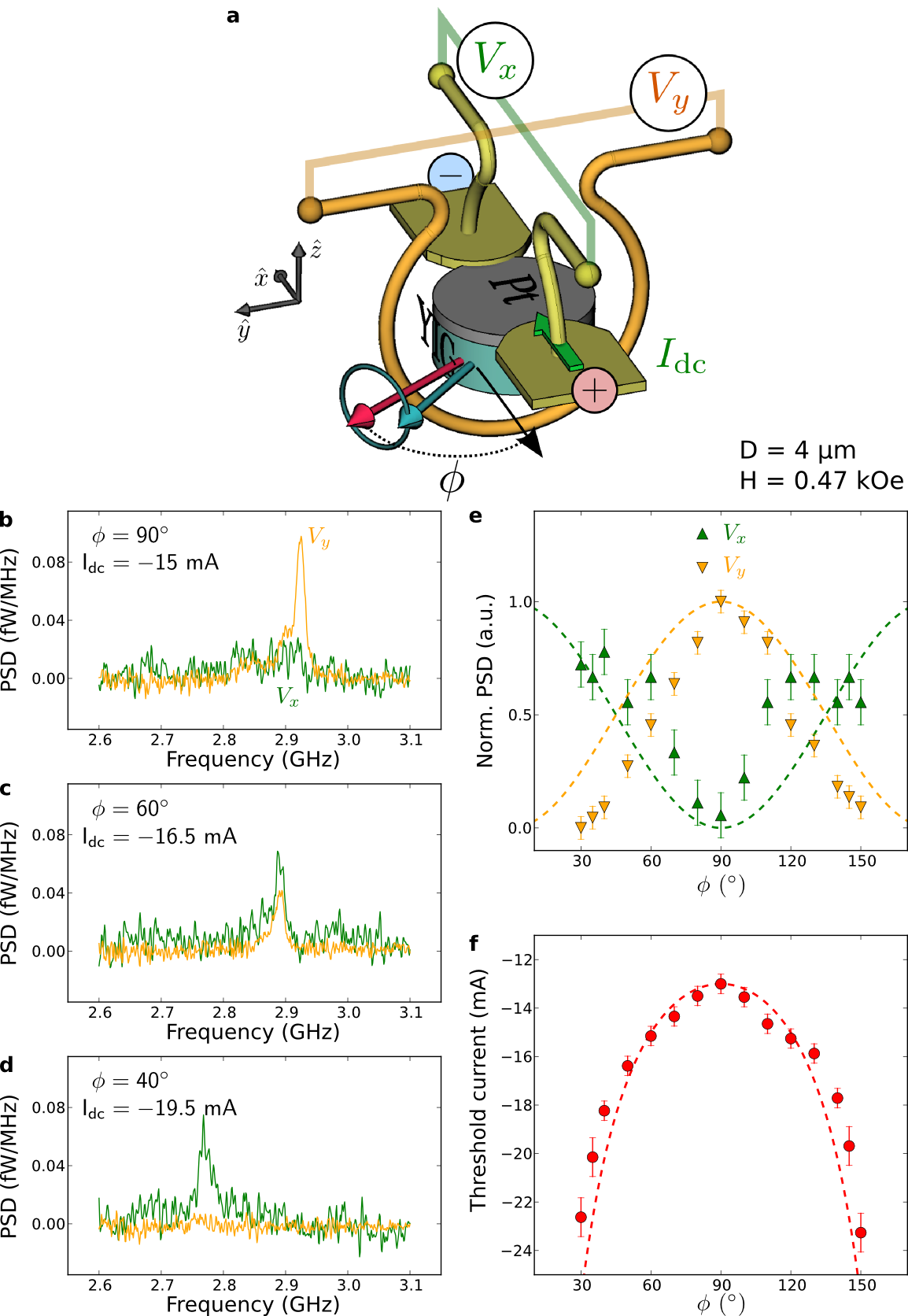}
  \caption{\textbf{Auto-oscillations as a function of the angle
      between the dc current and the bias field.}  \textbf{a}, Sketch
    of the sample and measurement configuration. The bias field $H$ is
    oriented at an angle $\phi$ with the dc current $I_\text{dc}$ in
    the Pt. The precession of the YIG magnetisation induces voltages
    $V_x$ in the antenna and $V_y$ across Pt, which are amplified and
    monitored by spectrum analysers. \textbf{b--d}, $V_x$ and $V_y$ at
    $H=0.47$~kOe for three different angles $\phi$. \textbf{e},
    Dependence of the normalized signals in both circuits and
    \textbf{f}, of the threshold current for auto-oscillations on
    $\phi$. Dashed lines show the expected angular dependences.}
  \label{FIG3}
\end{figure}

To gain further insight about the origin of the auto-oscillation
signal, we now monitor how the auto-oscillations of the 4~$\mu$m disc
evolve as the angle $\phi$ between the in-plane bias field fixed to
$H=0.47$~kOe and the dc current $I_\text{dc}$ is varied from
$30^\circ$ to $150^\circ$. The results are summarized in
Fig. 3. Pannels b--d show the auto-oscillation voltages detected in
the antenna ($V_y$) and across the Pt electrode ($V_x$). At
$\phi=90^\circ$, the auto-oscillation signal is only visible in the
$V_y$ channel. At $\phi=60^\circ$, both $V_x$ and $V_y$ channels
exhibit the auto-oscillation peak. At $\phi=40^\circ$, it almost
vanishes in $V_y$, while it slightly increases in $V_x$. The
normalized signals as a function of $\phi$ are plotted in Fig. 3e. The
Pt electrode and antenna loop being oriented perpendicularly to each
other (see Fig. 3a), the ac flux due to the precession of
magnetisation picked up by each of them respectively varies as
$\cos{\phi}$ and $\sin{\phi}$ (dashed lines in Fig. 3e).

% It is worth mentioning that due to ISHE in Pt, an ac spin pumping
% contribution \cite{hahn13a,weiler14,wei14} is also expected to be
% present in $V_x$. However, we estimate its amplitude to be at least
% ten times smaller than the inductive contribution in our samples
% \cite{hahn14a}, \textit{i.e.}, below the noise level.

More importantly, this study of angle dependence also allows us to
extract the threshold current for auto-oscillations as a function of
$\phi$. As $\phi$ deviates from the optimal orientation $90^\circ$,
the absolute value of the threshold current rapidly increases, see
Fig. 3f. In fact, the SOT acting on the oscillating part $\bf m$ of
the magnetisation scales as $\bf m \times \bf s \times \bf m \propto
\sin{\phi}$, where $\bf s$ is the spin polarisation of the dc spin
current produced by SHE in Pt at the YIG$|$Pt interface. Therefore,
the expected threshold current scales as $1/\sin{\phi}$, which is
plotted as a dashed line in Fig. 3f, in very good agreement with the
data.

% ****************************************************************
% Identification of the auto-oscillating mode and discussion

\begin{figure*}
  \includegraphics[width=17cm]{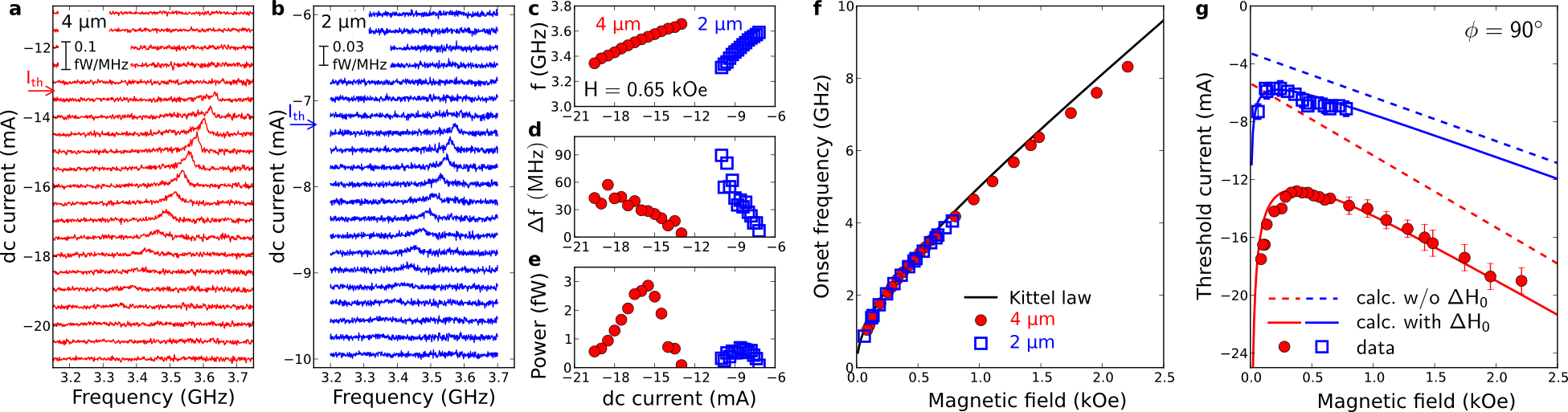}
  \caption{\textbf{Quantitative analysis of auto-oscillations in
      YIG$|$Pt microdiscs}. \textbf{a}, Inductive voltage $V_y$
    produced by auto-oscillations in the 4~$\mu$m and \textbf{b},
    2~$\mu$m YIG$|$Pt discs as a function of the dc current
    $I_\text{dc}$ in the Pt. The experimental configuration is the
    same as in Fig. 1a, with the bias field fixed to
    $H=0.65$~kOe. \textbf{c}, Auto-oscillation frequency, \textbf{d},
    linewidth and \textbf{e}, integrated power \textit{vs.}
    $I_\text{dc}$. \textbf{f}, Dependence of the onset frequency and
    \textbf{g}, of the threshold current on the applied field in both
    discs. Expectations taking into account only the homogeneous
    linewidth or the total linewidth are respectively shown by dashed
    and continuous lines.}
  \label{FIG4}
\end{figure*}

In summary, the results reported in Figs. 1 and 3 unambiguously
demonstrate that the auto-oscillations observed in our hybrid YIG$|$Pt
discs result from the action of SOT produced by $I_\text{dc}$. We have
also shown that they correspond to the reverse effect of the spin
pumping mechanism illustrated in Fig. 2d and its detection through
ISHE in Fig. 2b.

In the last part of this letter, we analyse quantitatively the main
features of auto-oscillations, which allows us to determine their
nature and to understand the role of quasi-degenerate SW modes in the
SOT driven dynamics.

For this, we compare the auto-oscillations observed in the 4~$\mu$m
and 2~$\mu$m microdiscs. Figs. 4a and 4b respectively present the
inductive signal $V_y$ detected in the antenna coupled to these two
discs as a function of $I_\text{dc}$. The configuration is the same as
depicted in Fig. 1a, with a slightly larger bias field set to
$H=0.65$~kOe. One can clearly see a peak appearing in the PSD close to
3.6~GHz in both cases, at a threshold current of about $-13.5$~mA in
the 4~$\mu$m disc and $-7.4$~mA in the 2~$\mu$m disc. These two values
correspond to a similar threshold current density in both samples of
$(4.4 \pm 0.2)\cdot10^{11}~\text{A.m}^{-2}$, in agreement with our
previous study \cite{hamadeh14b}. As the dc current is varied towards
more negative values, the peaks shift towards lower frequency
(Fig. 4c), at a rate which is twice faster in the smaller disc. This
frequency shift is mainly due to linear and quadratic contributions in
$I_\text{dc}$ of Oersted field and Joule heating, respectively
\cite{hamadeh14b} (from the Pt resistance, the maximal temperature
increase in both samples is estimated to be $+40^\circ$C).  At the
same time, the signal first rapidly increases in amplitude, reaches a
maximum, and then, more surprisingly, drops until it cannot be
detected anymore, as seen in Fig. 4e, which plots the integrated power
\textit{vs.} $I_\text{dc}$. The maximum of power measured in the
4~$\mu$m disc (2.9~fW) is four times larger than the one measured in
the 2~$\mu$m disc (0.7~fW), which is due to the inductive origin of
$V_y$. The latter can be estimated from geometrical considerations,
$V_y=\eta (\omega \mu_0 D t_\text{YIG} M_s \sin{\theta})/2$. Here
$\mu_0$ is the magnetic constant, $D$ the diameter of the disc and
$\theta$ the angle of uniform precession (the prefactor $\eta \simeq
0.1$ accounts for microwave losses and impedance mismatch in the
measured frequency range with our microwave circuit). For the same
$\theta$, the inductive voltage produced by the 4~$\mu$m disc is thus
twice larger than by one produced by the 2~$\mu$m disc, hence the
ratio four in power. Moreover, the maximal angle of precession reached
by auto-oscillations is found to be about $1^\circ$ in both microdiscs
\cite{hamadeh14b}. Finally, the disappearance of the signal as
$I_\text{dc}$ gets more negative is accompanied by a continuous
broadening of the linewidth, which increases from a few MHz to several
tens of MHz (Fig. 4d).  This rather large auto-oscillation linewidth
is also consistent with a small precession angle, \textit{i.e.}, a
small stored energy in the YIG oscillator \cite{kim08}.

By repeating the same analysis as a function of $H$, we can determine
the bias field dependence of the auto-oscillations in both
microdiscs. The onset frequency and threshold current at which
auto-oscillations start are plotted in Figs. 4f and 4g,
respectively. The onset frequency in the 4~$\mu$m and 2~$\mu$m
microdiscs is identical and closely follows the dispersion relation of
the main FMR mode plotted as a continuous line. The small redshift
towards lower frequency, which increases with the applied field, is
ascribed to the Joule heating and Oersted field induced by
$I_\text{dc}$ (the Kittel law in Figs. 2c and 4f is obtained at
$I_\text{dc}=0$~mA). We also note that the main FMR mode is the one
which couples the most efficiently to our inductive electrical
detection, because it is the most uniform. Hence, we conclude that the
detected auto-oscillations are due to the destabilisation of this mode
by SOT.

In order to reach auto-oscillations, the additional damping term due
to SOT has to compensate the natural relaxation rate $\Gamma_r$ in
YIG. Given the transparency $T$ of the YIG$|$Pt interface and the spin
Hall angle $\Theta_\text{SH}$ in Pt, this condition writes:
\begin{equation}
  T \Theta_\text{SH} \frac{\hbar}{2e} \frac{\gamma}{t_\text{YIG}M_s} \frac{I_\text{th}}{t_\text{Pt} D}=-\Gamma_r \, ,
  \label{threshold}
\end{equation}
where $t_\text{Pt} D$ is the section of the Pt layer. The homogeneous
contribution to $\Gamma_r$ is given by the Gilbert damping rate, which
for the in-plane geometry is:
\begin{equation}
  \Gamma_G=\alpha \gamma (H+2\pi M_s) \, .
  \label{Gilbert}
\end{equation}
We remind that this expression is obtained by converting the field
linewidth to frequency linewidth through $\Delta \omega = \Delta H
(\partial \omega/\partial H)$. If only the homogeneous contribution to
the linewidth is taken into account, the threshold current
$I_\text{th}$ is thus expected to depend linearly on $H$, as shown by
the dashed lines plotted in Fig. 4g using Eqs. \ref{threshold} and
\ref{Gilbert}, and the parameters listed in Table I (the only
adjustment made is for the 4~$\mu$m disc, where the calculated
$I_\text{th}$ has been reduced by 20\% in order to reproduce
asymptotically the experimental slope of $I_\text{th}$ \textit{vs.}
$H$). It qualitatively explains the dependence of $I_\text{th}$ at
large bias field in both microdiscs, but underestimates its value and
fails to reproduce the optimum observed at low bias field.

To understand this behaviour, the finite inhomogeneous contribution to
the linewidth $\Delta H_0$ measured in Fig. 2d should be considered as
well. In fact, this contribution dominates the full linewidth at low
bias field. In that case, the expression of the relaxation rate
writes:
\begin{equation}
  \Gamma_r=\Gamma_G + \gamma \frac{\Delta H_0}{2} \frac{H+2\pi M_s}{\sqrt{H(H+4\pi M_s)}} \, .
  \label{inhom}
\end{equation}
The form of the last term in Eq. \ref{inhom} is due to the Kittel
dispersion relation and is responsible for the existence of the
optimum in $I_\text{th}$ at $H \neq 0$. Using the value of $\Delta
H_0=0.7$~Oe extracted in Fig. 2d for the 2~$\mu$m disc in
Eq. \ref{inhom} in combination with Eq. \ref{threshold}, the
continuous blue line of Fig. 4g is calculated, in very good agreement
with the experimental data. To get such an agreement for the 4~$\mu$m
disc, $\Delta H_0$ has to be increased by 25\% compared to the value
determined in Fig. 2d. In this case, both the position of the optimum
(observed at $H\simeq0.3-0.5$~kOe) and the exact value of
$I_\text{th}$ are also well reproduced for the 4~$\mu$m disc, as shown
by the continuous red line in Fig. 4g.

Hence, it turns out that quasi-degenerate SW modes, which are
responsible for the inhomogeneous contribution to the linewidth,
strongly affect the exact value and detailed dependence \textit{vs.}
$H$ of $I_\text{th}$. In fact, it is the total linewidth that truly
quantifies the losses of a magnetic device regardless of the nature
and number of microscopic mechanisms involved. Even in structures with
micron-sized lateral dimensions, there still exist a few
quasi-degenerate SW modes as evidenced by the finite $\Delta H_0$
observed in Fig. 2d. Due to magnon-magnon scattering, they are
linearly coupled to the main FMR mode, which as a result has its
effective damping increased, along with the threshold current. The
presence of these SW modes is also known to play a crucial role in SOT
driven dynamics. The strongly non-equilibrium distribution of SWs
promoted by SOT in combination with nonlinear interactions between
modes can lead to mode competition, which might even prevent
auto-oscillations to start \cite{demidov11d}. We believe that the
observed behaviours of the integrated power (Fig. 4d) and linewidth
(Fig. 4e) \textit{vs.} $I_\text{dc}$ are reminiscent of the presence
of these quasi-degenerate SW modes. A meaningful interpretation of
these experimental results is that as the FMR mode starts to
auto-oscillate and to grow in amplitude as the dc current is increased
above the threshold, its coupling to other SW modes -- whose
amplitudes also grow due to SOT -- becomes larger, which makes the
flow of energy out of the FMR mode more efficient. This reduces the
inductive signal, as non-uniform SW modes are poorly coupled to our
inductive detection scheme. At the same time, it enhances the
auto-oscillation linewidth, which reflects this additional nonlinear
relaxation channel.

% ****************************************************************
% Conclusion

The smaller inhomogeneous linewidth in the 2~$\mu$m disc (Fig. 2d)
results in a field dependence of the threshold current closer to the
one expected for the purely homogeneous case (Fig. 4g). This indicates
that reducing further the lateral size of the microstructure will
allow to completly lift the quasi-degeneracy between spin-wave modes
\cite{hahn14}, as predicted by micromagnetic simulations, which show
that this is obtained for lateral sizes smaller than $1~\mu$m. This
could extend the stability of the auto-oscillation for the FMR mode,
and experimental techniques capable of detecting SWs in nanostructures
\cite{hamadeh14b,demidov11d} should be used to probe this
transition. Very importantly for the field of magnonics, it was
recently shown that this constraint on confinement could be relaxed in
one dimension such as to produce a propagation stripe
\cite{duan14}. Other strategies might consist in using specific
non-uniform SW modes or to engineer the SW spectrum using topological
singularities such as vortices, or bubbles, which could be most
relevant to design active magnonics computational circuits.

% ****************************************************************

\section*{Methods}

\textbf{Samples --} Details of the PLD growth of the YIG layer can be
found in ref.\cite{kelly13}. Its dynamical properties have been
determined by broadband FMR measurements. The transport parameters of
the 8~nm thick Pt layer deposited on top by magnetron sputtering have
been determined in a previous study \cite{rojas14}. The YIG$|$Pt
microdiscs are defined by e-beam lithography, as well as the
Au(80~nm)$|$Ti(20~nm) electrodes -- separated by 1~$\mu$m from each
other -- which contact them. This electrical circuit is insulated by a
300~nm thick SiO$_2$ layer, and a broadband microwave antenna made of
250~nm thick Au with a 5~$\mu$m wide constriction is defined on top of
each disc by optical lithography.

\textbf{Measurements --} The samples are mounted between the poles of
an electromagnet which can be rotated to vary the angle $\phi$ shown
in Fig. 3a. Two 50~$\Omega$ matched picoprobes are used to connect to
the microwave antenna and to the electrodes which contact the Pt
layer. The latter are connected to a dc current source through a
bias-tee. To perform ISHE-detected FMR measurements, a microwave
synthesizer is connected to the microwave antenna, and the output
power is turned on and off at a modulation frequency of 9~kHz. The
voltage across Pt is measured by a lock-in after a low-noise
preamplifier (gain 100). For the detection of auto-oscillations,
high-frequency low-noise amplifiers are used (gain 33~dB to 39~dB,
depending on the frequency range). Two spectrum analysers
simultaneously monitor in the frequency domain the voltages $V_x$ and
$V_y$ across the Pt layer and in the microwave antenna, respectively
(Fig. 3a). The resolution bandwidth employed in the measurements is
set to 1~MHz.

\begin{acknowledgments}
  We acknowledge E. Jacquet, R. Lebourgeois and A. H. Molpeceres for
  their contribution to sample growth, and M. Viret and A. Fert for
  fruitful discussion. This research was partially supported by the
  ANR Grant Trinidad (ASTRID 2012 program). V. V. N. acknowledges
  support from the program CMIRA'Pro of the region
  Rh\^one-Alpes. S. O. D. and V. V. N. acknowledge respectively
  support from the Russian programs Megagrant No. 2013-220-04-329 and
  Competitive Growth of KFU.
\end{acknowledgments}

\end{document}